\documentstyle[preprint,epsf,aps]{revtex}
\tightenlines

\draft

\begin{document}
\title{Superdecay of excited atom placed near ''Left-Handed'' sphere}
\author{V.V.Klimov}
\address{P.N.Lebedev Physical Institute, Russian Academy of Sciences, 53 Leninsky\\
Prospect, Moscow, 117924, Russia }
\date{\today}
\maketitle

\begin{abstract}
The influence of a microsphere having a simultaneously negative permittivity
and permeability (Left-Handed Sphere) on the decay rate of the allowed (~$%
E_{1}$~) and forbidden (~$M_{1}$~) transitions is considered. It is found,
that in contrast to usual ( Right-Handed) materials it is possible to
increase the decay rate by several orders of magnitude even in the case of a
sphere of small radii. This enhancement is due to the fact that there exists
a new type of resonance modes (LH surface modes). The analytical expressions
for resonance frequencies and associated quality factors of the new modes
are found.\newline
\end{abstract}


\newpage In paper \cite{Veselago} , published in 1967, it was shown , that
hypothetical media with a simultaneously negative permittivity and
permeability have a number of unusual properties. Here, in particular, the
phase and group velocities of light propagation have opposite directions of
propagation . The detailed analysis of this effect was carried out in \cite
{Smith1}. These materials were referred to as Left-Handed (LH) materials in
contrast to the usual Right-Handed (RH) materials. The Doppler effect has an
unusual form in such media. For example, when the distance between the
source and receiver diminishes , the frequency picked by the receiver will
be less than the source frequency as compared to usual media. The
Vavilov-Cherenkov effect is also drastically modified: the energy flow from
the particle moving faster than the phase velocity in LH medium propagates
in backward direction, while in RH medium we have the forward radiation.

The influence of the interface between LH and RH media on the radiation
propagation is described by the reflection and the refraction laws, which
differ substantially from the laws specific for the usual media. For
example, within such an interface the refraction angle becomes negative, and
the reflection beam can not exist at all \cite{Veselago}.

Until the present time the existence of LH media remained unproved. This
fact has not stimulated further investigation of such media. With the recent
experimental demonstration of principal possibility of creation of LH media
on the base of composite materials \cite{Smith2} , it is now relevant to
investigate in more detail the influence of LH media on more complicated
processes, such as spontaneous emission.

Purcell \cite{Purcell} was the first who pointed out that spontaneous decay
rates can be modified in a resonant cavity. This idea received a strong
proof in the experiments with atoms in microwave cavities \cite{Goy,Jhe} and
with $Eu^{3+}$ ions in liquid microdroplets \cite{Lin}. In \cite
{Bykov,Yablonovich,John} it was shown, that spontaneous emission can be
substantially inhibited in 3-dimensional periodic dielectric structures
(photonic crystals). The influence of a semi-infinite active media on
spontaneous emission was investigated in \cite{Kocharovsky}. This
investigation showed that the instability of the ground state of an atom
arises, and the radiation reaction force changes its sign in comparison with
the usual (passive) media.

In the present paper the processes of spontaneous emission of the atom placed
near the body (sphere) made of LH material and placed in the RH medium will
be considered, and it will be shown that these processes drastically differ
from those near usual RH body. The geometry of the problem is shown in Fig.1.

When neglecting the losses one can obtain, within the classical \cite{Chew}
and quantum \cite{Klimov1,Klimov2} perturbative approaches, the expressions
for spontaneous decay rates for an excited atom placed near the sphere of
the radius $a$ with arbitrary permittivity and permeability. These
approaches are valid when the decay rate $\gamma $ is small in comparison
with radiation frequency $\omega$, $\gamma\ll\omega$. Besides, this
condition allows us to neglect the frequency dependence of $\epsilon $ and $%
\mu $. For the electric dipole (${E_{1}}$) transitions the decay rates
normalized to the outer space values have the following form

\begin{equation}
\left( \frac{\gamma }{\gamma _{0}}\right) _{rad}^{E_{1}}=\frac{3}{2}%
\sum\limits_{n=1}^{\infty }n\left( n+1\right) \left( 2n+1\right) \left( 
\frac{\left| j_{n}\left( k_{2}r\right) -q_{n}h_{n}^{\left( 1\right) }\left(
k_{2}r\right) \right| }{k_{2}r}\right) ^{2}
\end{equation}

for the radial orientation of the electric dipole momentum and

\begin{equation}
\left( \frac{\gamma }{\gamma _{0}}\right) _{tan}^{E_{1}}=\frac{3}{2}%
\sum\limits_{n=1}^{\infty }\left( n+1/2\right) \left[ \left| j_{n}\left(
z\right) -p_{n}h_{n}^{\left( 1\right) }\left( z\right) \right| ^{2}+\left| 
\frac{{\ d}\left( {z}\left( j_{n}\left( z\right) -q_{n}h_{n}^{\left(
1\right) }\left( z\right) \right) \right) }{zdz}\right| ^{2}\right]
_{z=k_{2}r}
\end{equation}

for the tangential orientations.

One can show that decay rates of the magnetic dipole (${M_{1}}$) transitions
can be derived from (1,2) by the interchange of $\varepsilon $ and $\mu $ 
\cite{Klimov2}. As a result we have

\begin{equation}
\left( \frac{\gamma }{\gamma _{0}}\right) _{rad}^{M_{1}}=\frac{3}{2}%
\sum\limits_{n=1}^{\infty }n\left( n+1\right) \left( 2n+1\right) \left( 
\frac{\left| j_{n}\left( k_{2}r\right) -p_{n}h_{n}^{\left( 1\right) }\left(
k_{2}r\right) \right| }{k_{2}r}\right) ^{2}
\end{equation}

for the radial orientation of the magnetic dipole momentum and

\begin{equation}
\left( \frac{\gamma }{\gamma _{0}}\right) _{tan}^{M_{1}}=\frac{3}{2}%
\sum\limits_{n=1}^{\infty }\left( n+1/2\right) \left[ \left| j_{n}\left(
z\right) -q_{n}h_{n}^{\left( 1\right) }\left( z\right) \right| ^{2}+\left| 
\frac{d\left( z\left( j_{n}\left( z\right) -p_{n}h_{n}^{\left( 1\right)
}\left( z\right) \right) \right) }{zdz}\right| ^{2}\right] _{z=k_{2}r}
\end{equation}

for the tangential orientations.

In (1-4) $j_{n}$ and $h_{n}^{\left( 1\right) }$ are the spherical Bessel and
Hankel functions \cite{Abramowitz}; $z_{1,2}=k_{1,2}a$ , $k_{1,2}=\sqrt{%
\epsilon _{1,2}\mu _{1,2}}\omega /c$ , $q_{n}$ and $p_{n}$, the Mie
coefficients of the spherical wave reflection. For TM reflection
coefficients, $q_{n}$ , we have the expression

\begin{equation}
q_{n}=\frac{\displaystyle\varepsilon _{1}\frac{d}{dz_{2}}\left[
z_{2}j_{n}\left( z_{2}\right) \right] j_{n}\left( z_{1}\right) -\varepsilon
_{2}\frac{d}{dz_{1}}\left[ z_{1}j_{n}\left( z_{1}\right) \right] j_{n}\left(
z_{2}\right) }{\displaystyle\varepsilon _{1}\frac{d}{dz_{2}}\left[
z_{2}h_{n}^{\left( 1\right) }\left( z_{2}\right) \right] j_{n}\left(
z_{1}\right) -\varepsilon _{2}\frac{d}{dz_{1}}\left[ z_{1}j_{n}\left(
z_{1}\right) \right] h_{n}^{\left( 1\right) }\left( z_{2}\right) }
\end{equation}

The expressions for TE reflection coefficients, $p_{n}$ are derived from $%
q_{n}$ by the substitution $\epsilon \Leftrightarrow \mu $.

It is important that the expressions (1-5) have been found in the course of
direct solution of Maxwell equations. This procedure does not depend on the
signs of permittivity and permeability. Moreover, these expressions have no
branching point. That is why we do not need to analyze different sheets of
complex plane. In other words, the sign of the refraction index ( negative
for LH media \cite{Veselago,Smith1}) is inessential within the approach,
which uses the standing waves to describe the field inside the sphere. Thus,
the expressions (1-5) are valid for any media, including the RH and LH ones.

From physical point of view any body influences spontaneous emission of an
atom through the reflected electromagnetic waves modifying the dynamics of
the electron. One can expect that the properties of the reflected field will
be quit different for the RH and LH bodies, since the refraction and
reflection laws are different for them\cite{Veselago}. In Figs.2,3 the ray
trajectories from the dipole source near the surface of RH and LH spheres are
shown. From this picture one can see that the character of penetration of
the dipole field to the sphere, and, hence, the character of the sphere
excitation is different for different materials. The RH sphere defocuses the
rays (Fig.2), while the LH one focuses them(Fig.3).

To analyze the electromagnetic processes in a microsphere quantitatively one
should consider the Mie reflection coefficients (5). If the coefficients are
close to 1, the reflection is effective and a strong influence of the
reflected field on the emission process will take place.

In the case of RH sphere, Mie coefficients are close to 1 if the resonance
modes ( whispering gallery modes) are excited in the microsphere. The
whispering gallery modes can have very high quality factors \cite
{Braginsky1,Braginsky2,Collot}, and a microlaser with superlow threshold of
generation has been developed on the base of silica microspheres \cite
{Sandoghdar}.

One can find the approximate conditions of the whispering gallery modes
appearance if one requires that the integer number of the wavelength fits in
the sphere perimeter

\begin{equation}
2\pi a=n\frac{\lambda _{vac}}{\sqrt{\epsilon _{1}\mu _{1}}},n\gg 1
\end{equation}

More exact values for the resonant frequencies and associated quality
factors one can find by making the denominators of $q_{n}$ and $p_{n}$ equal
to zero \cite{Klimov3}. It is important that the high quality whispering
gallery modes can exist only in a sphere with the radius, larger than the
wavelength.

The excitation of the whispering gallery modes is possible in the LH sphere
too. However, these modes are of no importance in the case of LH spheres
with the radii comparable to the wavelength or less. The surface LH modes,
having no analogy in RH case, play the main role in this domain. The
dependence of TE reflection coefficients on the radius \ $a$ of a sphere made of
RH and LH material is shown in Fig.4. One can see that the usual whispering
gallery modes are typical for the sphere of large radius, while the new LH
surface mode appears in the case of a small LH sphere. Only the surface TE
mode appears in the parameter domain under consideration .

It is easily seen that these modes have no maxima inside the sphere, and it
is naturally to refer to this modes as the surface LH modes. These modes
have no simple ray interpretation because they can exist even in the case of
LH sphere with the radius small in comparison to the wavelength.

One can find the approximate values for complex resonant frequencies ( or
for resonant radii corresponding to the given frequency) by making the
denominators of $q_{n}$ and $p_{n}$ equal to zero, and using small sphere
approximation $\left( 2\pi a/\lambda _{vac} \ll n\right) $ .

For TM modes the resonant wavelength is defined by the real part of complex
root $z_{TM}$

\begin{equation}
\Re\left( z_{TM} \right) = \frac{2\pi a}{\lambda _{vac}}=\sqrt{\frac{%
\epsilon _{2}+n\left( \epsilon _{1}+\epsilon _{2}\right) }{\epsilon
_{1}\epsilon _{2}\left( \mu _{1}/\left( 2n+3\right) +\mu _{2}/\left(
2n-1\right) \right) }}
\end{equation}

In the case of TE modes we have

\begin{equation}
\Re\left( z_{TE} \right) =\frac{2\pi a}{\lambda _{vac}}=\sqrt{\frac{\mu
_{2}+n\left( \mu _{1}+\mu _{2}\right) }{\mu _{1}\mu _{2}\left( \epsilon
_{1}/\left( 2n+3\right) +\epsilon _{2}/\left( 2n-1\right) \right) }}
\end{equation}

For quality factors of these modes we have the following expressions;

\begin{equation}
Q_{TM}^{-1}=\frac{2\Im (z_{TM})}{\Re (z_{TM})}=\left| \frac{\mu _{2}}{%
\displaystyle\left( \frac{\mu _{1}}{2n+3}+\frac{\mu _{2}}{2n-1}\right) }%
\frac{\left( \sqrt{\varepsilon _{2}\mu _{2}}z_{TM}\right) ^{2n-1}}{\left(\left(
2n-1\right) !!\right)^{2}}\right|
\end{equation}

\begin{equation}
Q_{TE}^{-1}=\frac{2\Im (z_{TE})}{\Re (z_{TE})}=\left| \frac{\epsilon _{2}}{%
\displaystyle\left( \frac{\epsilon _{1}}{2n+3}+\frac{\epsilon _{2}}{2n-1}%
\right) }\frac{\left( \sqrt{\varepsilon _{2}\mu _{2}}z_{TE}\right) ^{2n-1}}
{\left(\left(2n-1\right)!!\right)^{2}}\right|
\end{equation}

From the analysis given in (7-10) and the derivation procedure one can see,
that such surface modes may appear only in LH sphere, and their quality
factor can be very large. It should be noted that these modes as, distinct
from the whispering gallery modes, can be excited in the LH sphere of an
arbitrary small radius. In this case the fulfillment of condition $\epsilon
_{2}+n\left( \epsilon _{1}+\epsilon _{2}\right) \approx 0$ for TM modes and $%
\mu _{2}+n\left( \mu _{1}+\mu _{2}\right) \approx 0$ for TE modes is
necessary. Besides, in the LH sphere with given $\epsilon $ and $\mu $ the
surface resonances occur only for restricted angular momentum numbers, $%
n<n_{\max}$, where

\begin{equation}
n_{\max ,TM}\approx -\frac{\epsilon _{2}}{\epsilon _{1}+\epsilon _{2}}
\end{equation}

\begin{equation}
n_{\max ,TE}\approx -\frac{\mu _{2}}{\mu _{1}+\mu _{2}}
\end{equation}

In contrast to the whispering gallery modes, for the surface LH resonances,
each angular momentum number $n$ is associated with the only one LH surface
mode.

It is very important that in the LH sphere the imaginary part of complex
frequency $\omega _{res}$ has positive value. This contrasts to the case of
the RH sphere, where the imaginary part takes negative values. This means
that the definition of the quality factor (Q) through the ratio of the real
to the imaginary part of complex resonant frequency $\omega _{res}$ differs
by the sign from the case of the usual RH media. The definition of the
quality factor Q for an arbitrary medium should be written as follows

\begin{equation}
Q=-\beta \frac{2\Im\left( \omega _{res} \right)}{\Re\left( \omega _{res}
\right)}
\end{equation}

where $\beta =+1$ or $-1$ for the RH or LH media respectively. The change of
the sign in one or other definition is a characteristic feature of the LH
media. In particular, such situation occurs for the definition of the
Doppler and Vavilov-Cherenkov effects \cite{Veselago}.

The occurrence of a new type of high quality oscillations in the LH sphere
will strongly modify the spontaneous decay rates of an atom located near its
surface. The dependence of $E_{1}$ and $M_{1}$ decay rates of an atom placed
in close vicinity to the sphere surface (~$r\rightarrow a$~) on the sphere
radius $a$ is shown in Figs.5,6. The figures shows that the decay rates
can be enhanced by a factor of $10^{5}$ and more. This super enhancement is
due to the LH surface modes described above. The well pronounced peaks,
connected with LH surface modes with different angular numbers $n$ , can be
seen in detail on the insets. \ On the other hand, it is clearly seen from
the figures, that the whispering gallery modes result only in a rather small
variations of the decay rates. Of course, taking into account the losses,
the frequency dependence of $\epsilon $ and $\mu $ \ and other imperfections
can reduce the decay rates, but we believe that final effect will be large
enough.

The influence of the LH sphere on the rates of the forbidden ($M_{1}$)
decays is of great importance, since, as a result of super enhancement, the $%
M_{1}$ decay rate can approach to the decay rate of the allowed ($E_{1}$)
transitions.

In principle, the super enhancement effect can be used to create a high
precision photon source (photon on demand) by putting LH microsphere closer
or further to the path of the excited atoms: the atom passing near the resonant
LH microsphere will emit the photon with the probability equal to 1. In its
turn, the manipulation of the microsphere can be realized by an external
laser beam \cite{Ashkin}. In doing so, of course, one should take into
account the fact, that the force of the laser light on LH spheres can have the
peculiarities\cite{Veselago}, which deserve special consideration.

In conclusion, the spontaneous decay of an atom placed near the sphere made
of the material having simultaneously negative permittivity and permeability
(Left-Handed sphere) is considered. We have found that there exist the
surface resonant modes even in the case of an arbitrary small Left-Handed
sphere. These modes have no analogy with the modes in the usual
(Right-handed) sphere and result in super enhancement of decay rates of
allowed ($E_{1}$) and forbidden ($M_{1}$) transitions.

The author thanks the Russian Basic Research Foundation and the Center
'Integration' for financial support of this work.

\newpage 
\begin{figure}
\epsfxsize 3 in
\centerline{\epsfbox{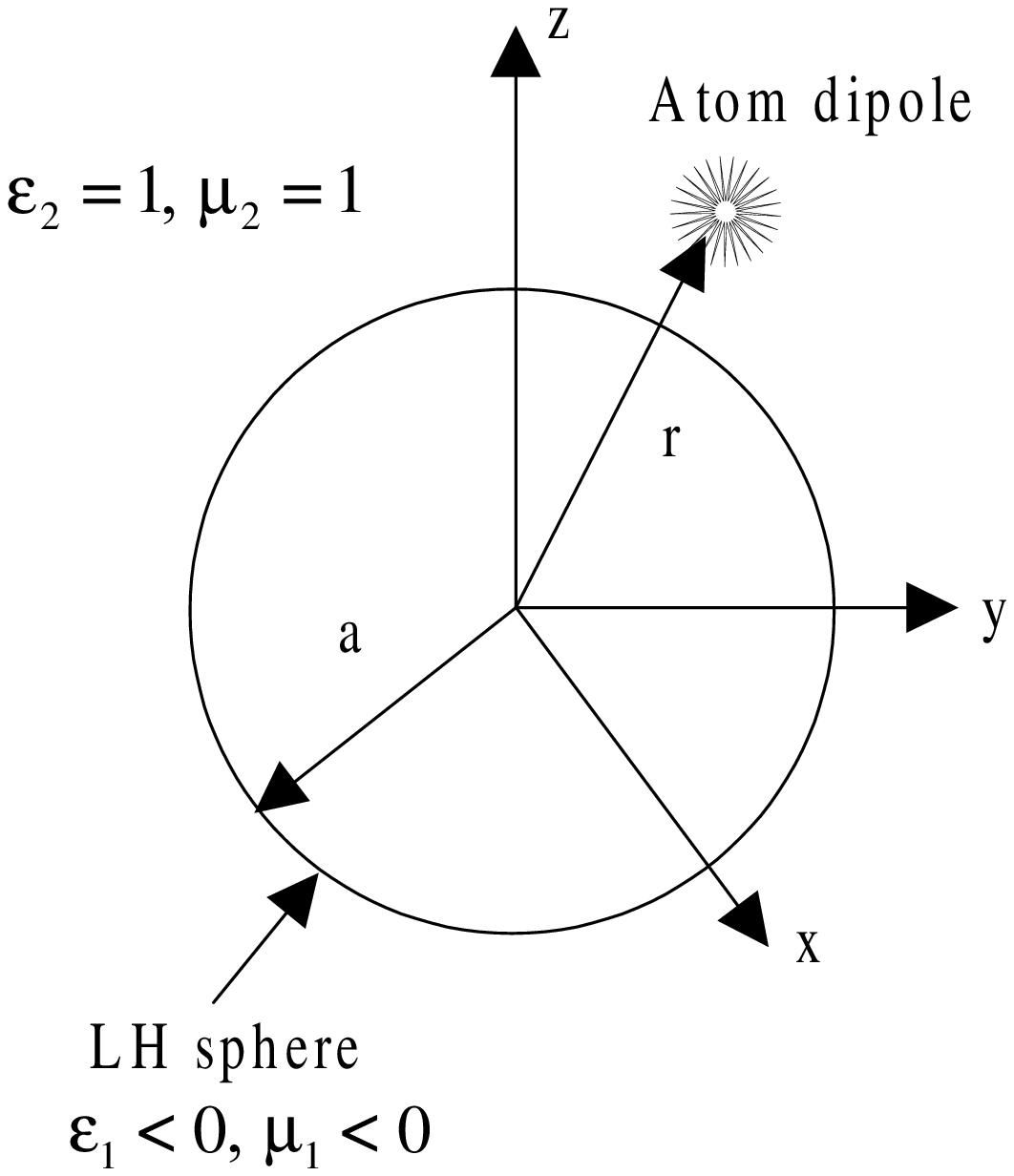}}
\medskip
\caption{Geometry of the problem}
\end{figure}

\begin{figure}
\epsfxsize 3 in
\centerline{\epsfbox{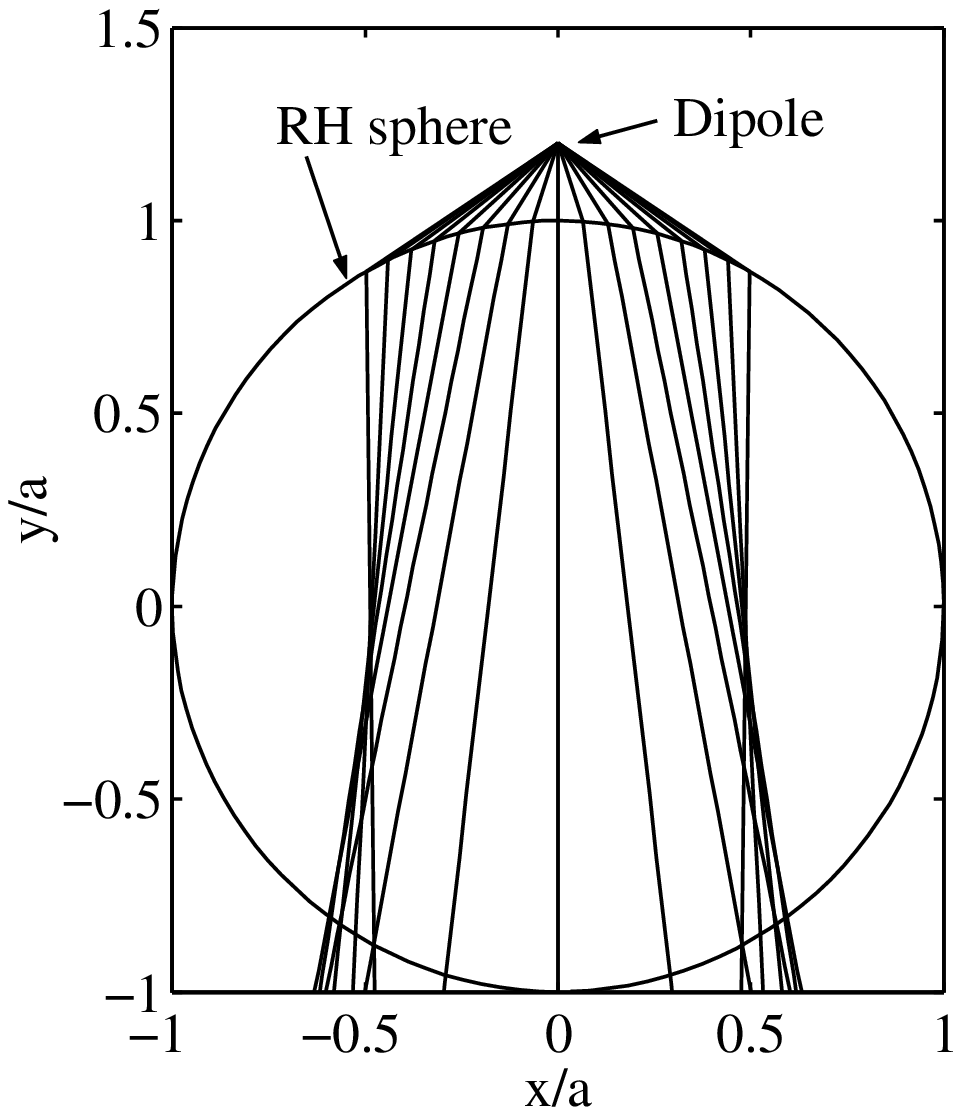}}
\medskip
\caption{Ray picture of excitation of the RH sphere with \ $\protect\epsilon %
_{1}=4$, \ $\protect\mu _{1}=1.05$}
\end{figure}
\newpage
\begin{figure}
\epsfxsize 3 in
\centerline{\epsfbox{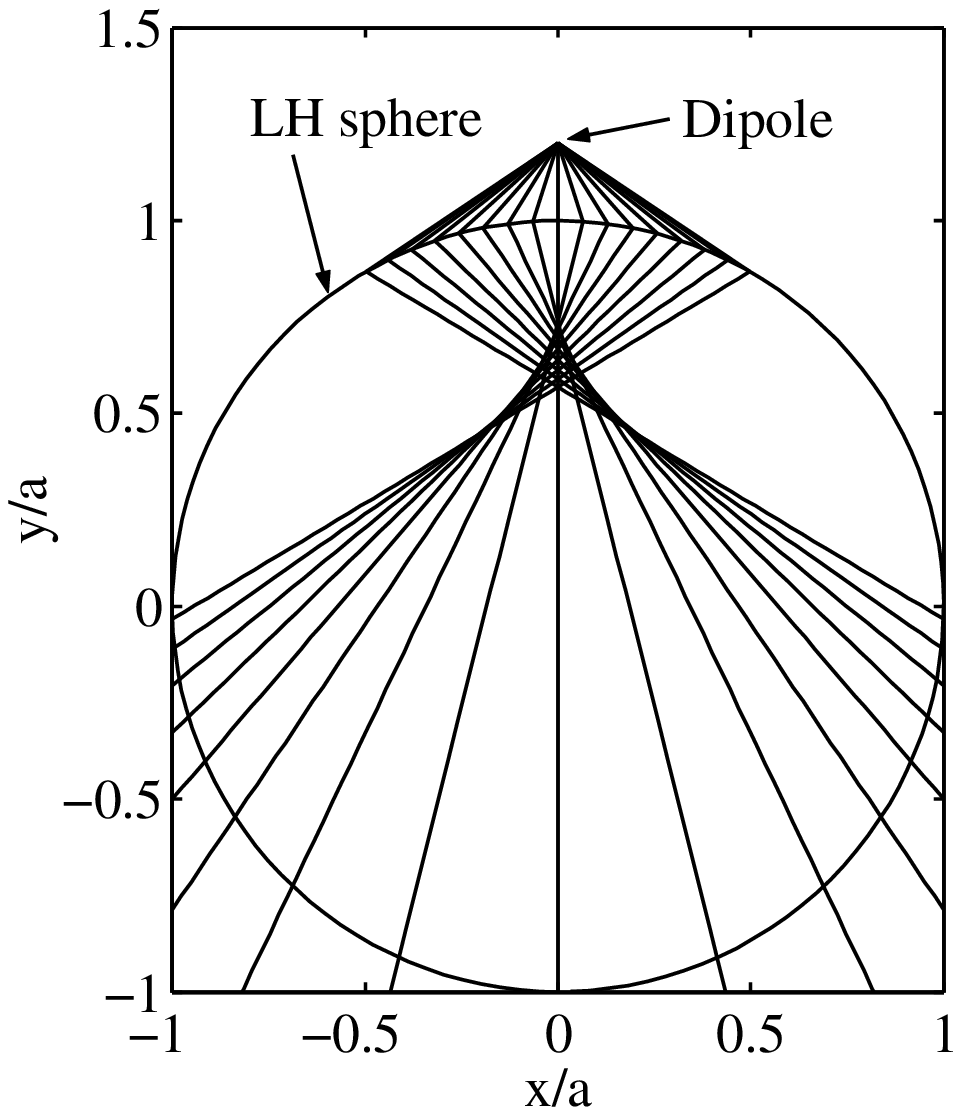}}
\medskip
\caption{Ray picture of excitation of the LH sphere with \ $\protect\epsilon %
_{1}=-4$, \ $\protect\mu _{1}=-1.05$ }
\end{figure}

\begin{figure}
\epsfxsize 3 in
\centerline{\epsfbox{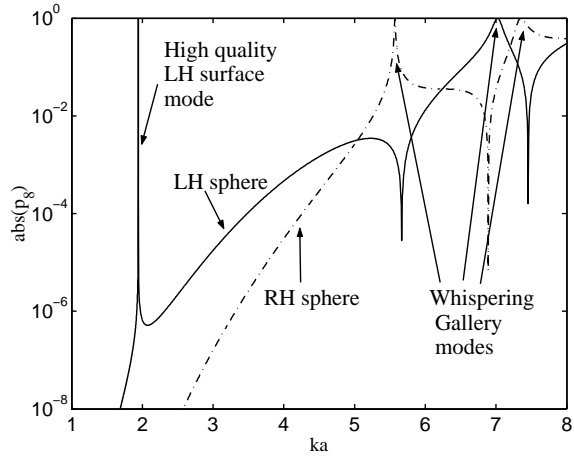}}
\medskip
\caption{The dependence of TE reflection coefficient $p_{8}$ on the sphere
radius $ka$ for RH sphere with $\protect\epsilon _{1}=4$, \ $\protect\mu %
_{1}=1.05$ (dash-dot line), and LH sphere with $\protect\epsilon _{1}=-4$, \ 
$\protect\mu _{1}=-1.05$ (solid line). }
\end{figure}
\newpage
\begin{figure}
\epsfxsize 3 in
\centerline{\epsfbox{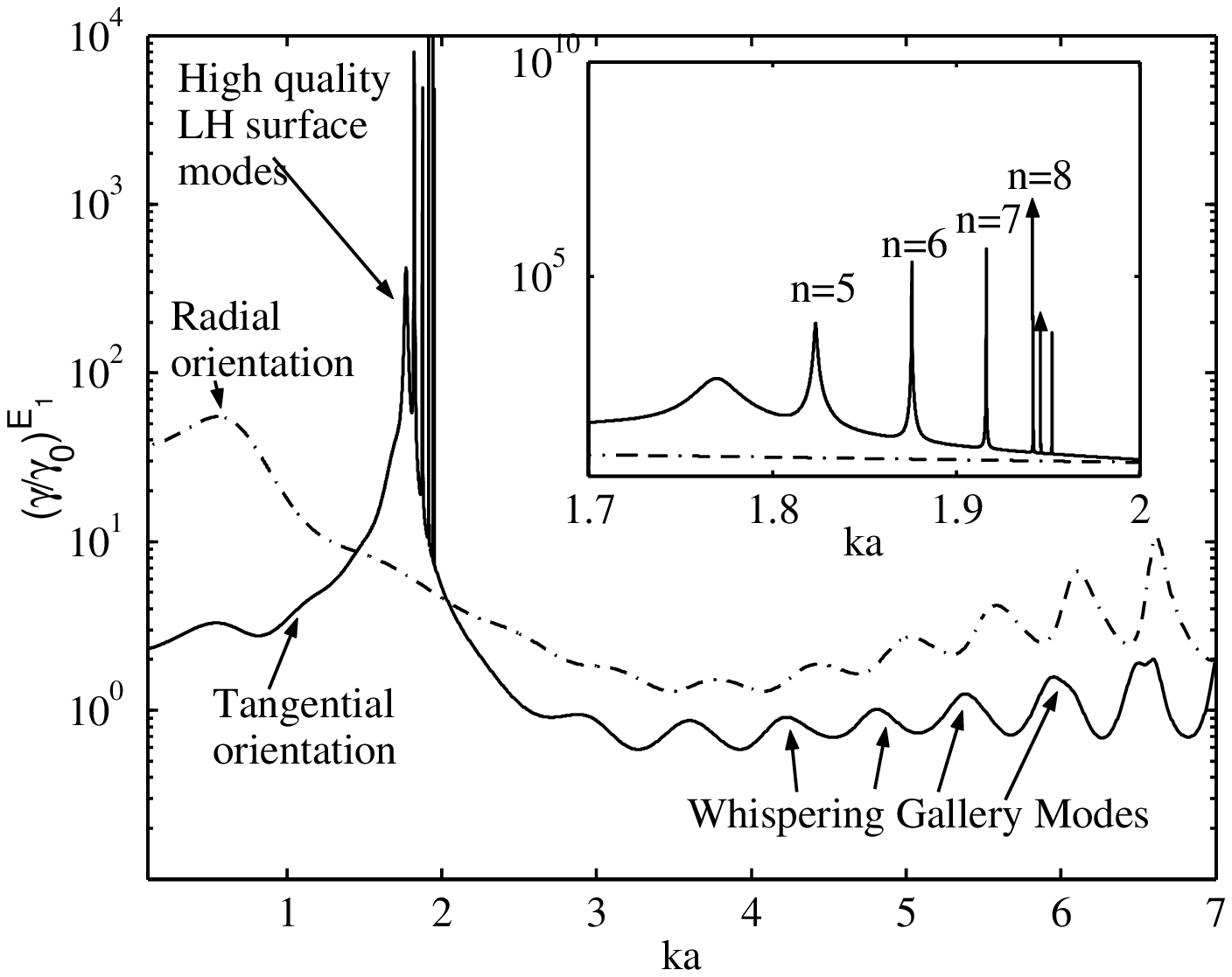}}
\medskip
\caption{The dependence of $E_{1}$ decay rates on the sphere radius $ka$
for an atom placed in close vicinity to the surface of LH sphere $%
(r\rightarrow a)$ with $\protect\epsilon _{1}=-4$, \ $\protect\mu _{1}=-1.05$%
}
\end{figure}

\begin{figure}
\epsfxsize 3 in
\centerline{\epsfbox{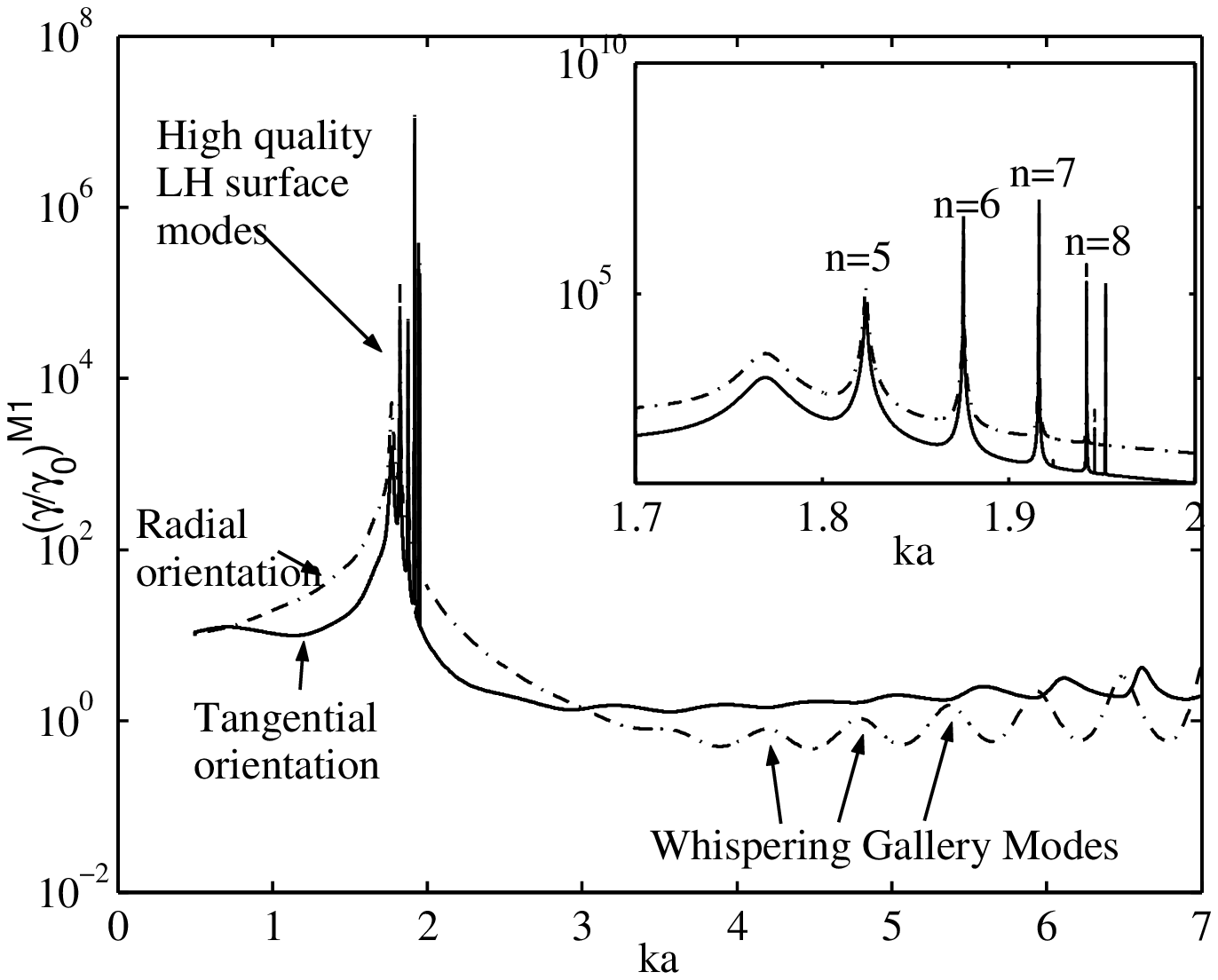}}
\medskip
\caption{The dependence of $M_{1}$ decay rates on the sphere radius $ka$
for an atom placed in close vicinity to the surface of LH sphere $%
(r\rightarrow a)$ with $\protect\epsilon _{1}=-4$, \ $\protect\mu _{1}=-1.05$%
}
\end{figure}

\end{document}